\documentclass[twocolumn,amsart,english,showpacs,preprintnumbers,amsmath,amssymb,floatfix]{revtex4-1}

\usepackage{tikz,xcolor}
\usepackage[colorlinks = true,
linkcolor = blue,
urlcolor  = blue,
citecolor = blue,
anchorcolor = blue]{hyperref}
\usepackage{lipsum}
\definecolor{lime}{HTML}{A6CE39}
\DeclareRobustCommand{\orcidicon}{%
    \begin{tikzpicture}
        \draw[lime, fill=lime] (0,0)
        circle [radius=0.16]
        node[white] {{\fontfamily{qag}\selectfont \tiny ID}};
        \draw[white, fill=white] (-0.0625,0.095)
        circle [radius=0.007];
    \end{tikzpicture}
    \hspace{-2mm}
}

\foreach \x in {A, ..., Z}{%
    \expandafter\xdef\csname orcid\x\endcsname{\noexpand\href{https://orcid.org/\csname orcidauthor\x\endcsname}{\noexpand\orcidicon}}
}

\usepackage[T1]{fontenc}
\usepackage[latin9]{inputenc}
\usepackage{color}
\usepackage{array}
\usepackage{amstext}
\usepackage{graphicx}
\usepackage{esint}
\usepackage{rotating}
\usepackage{appendix}
\usepackage{float}

\usepackage{hyperref}
\usepackage{xcolor}
\usepackage{soul}

\makeatletter

\@ifundefined{textcolor}{}
{%
    \definecolor{BLACK}{gray}{0}
    \definecolor{WHITE}{gray}{1}
    \definecolor{RED}{rgb}{1,0,0}
    \definecolor{GREEN}{rgb}{0,1,0}
    \definecolor{BLUE}{rgb}{0,0,1}
    \definecolor{CYAN}{cmyk}{1,0,0,0}
    \definecolor{MAGENTA}{cmyk}{0,1,0,0}
    \definecolor{YELLOW}{cmyk}{0,0,1,0}
}

\@ifundefined{definecolor}
{\usepackage{color}}{}
\@ifundefined{definecolor}
{\usepackage{color}}{}
\makeatother
\usepackage{babel}

\begin{document}

 \title{Role of universal function of the nuclear proximity potential: A systematic study on the alpha-decay of heavy/super-heavy nuclei and $\alpha$-induced reactions}

  \author{S. Mohammadi$^{1}${}}  

 \author{R. Gharaei$^{2}$\orcidA{}}
    \thanks{Electronic address}
    \email{rezagharaei@um.ac.ir}

    \author{S. A. Alavi$^{1}${}}
     
    \affiliation {$^{1}$Department of Physics, Sciences Faculty, Hakim Sabzevari University, P. O. Box 397, Sabzevar, Iran
\\$^{2}$Department of Physics, Faculty of Science, Ferdowsi University of Mashhad, P.O. Box 91775-1436, Mashhad, Iran}

    \date{\today}

%
%
\begin{abstract}\label{abstract}
The idea of the universal function is fundamental advantage of proximity potential model. In recent years, several formulae for this function have been given in the framework of the original proximity potential 1977 (Prox. 77). We present a systematic study of the role of universal function of the proximity potential model on $\alpha$-decay process for 250 ground-state to ground-state transitions using the WKB approximation. In order to realize this goal, five universal functions proposed in the proximity models gp 77, Guo 2013, Ngo 80, Zhang 2013 and Prox. 2010 have been incorporated into the formalism of Prox. 77. The obtained results show that the radial behavior of universal function affects the penetration probability of the $\alpha$ particle. The theoretical $\alpha$-decay half-lives are calculated and compared with the corresponding experimental data. It is that the Prox. 77 with Guo 2013 universal function provides the best fit to the available data (with standard deviation $\sigma = 0.49$). The Geiger-Nuttall (GN) plots for various isotopic groups have been investigated using this modified form of the proximity potential model. The role of the different universal functions in the $\alpha$-decay half-lives of super-heavy nuclei (SHN) with $Z=104-118$ are also studied. It is shown that the experimental half-lives in the super-heavy mass region are described well using the Prox. 77 with Zhang 2013 universal function. In this paper, the validity of the original and modified forms of the proximity 77 potential  is also examined for complete fusion reactions between $\alpha$-particle and 10 different target nuclei. Our results show that the measured $\alpha$-induced fusion reaction cross-sections can be well reproduced using the Prox. 77 with Zhang 2013 universal function for the reactions involving light and medium nuclei. Whereas, the Prox. 77 with Guo 2013 universal function model demonstrates a reliable agreement with the experimental data at sub-barrier energies for heavier systems. 
\\
\\
PACS number(s): 23.60.+e, 21.10.Tg, 24.10.-i
\end{abstract}
\maketitle

%
%
\section{INTRODUCTION}\label{int}

Alpha decay plays a fundamental role in nuclear physics and has been extensively studied in various theoretical and experimental frameworks \cite{sun,den,poen,yu}. It can provide a powerful tool to reveal some nuclear structure information such as the nuclear deformation, the properties of the ground state, the shell effects, energy levels, the effective nuclear interaction and so on \cite{seif,khu,car,del,shal,ren,qi,ren1,hod,zha}. Moreover, alpha decay is an accurate way to identify the new synthesized super-heavy nuclei \cite{mo,og}. Hence, $\alpha$-decay is the one of the most important decay modes of heavy and super-heavy nuclei. This decay was identified as $^4$He emission by the UK Professor Ernest Rutherford \cite{rut}. Gamow \cite{gam} and Condon and Gurney \cite{con} explained the alpha-decay process as a quantum mechanical tunneling effect. It is assumed that the alpha particle interacts with a daughter nucleus by tunneling through the Coulomb barrier which is formed in front of them \cite{gol}. So, the height and position of the potential barrier as well as its shape play a crucial role in the calculation of $\alpha$-decay half-lives. Accurate prediction of the $\alpha$-decay half-life requires a thorough understanding of the potential barrier. The total potential is composed of nuclear, Coulomb and centrifugal potentials as a function of the separation distance between two interacting nuclei. Due to the incomplete understanding of the nuclear potential, the accurate calculation of the interaction potential has remained a challenging task. Nevertheless, there are many approaches to determine the nuclear potential between the $\alpha$-particle and the daughter nucleus, such as the Coulomb and proximity potential model \cite{san,san1,deng}, the density-dependent cluster model with a double-folding integral of the renormalized M3Y nucleon-nucleon potential \cite{xu,xu1,xu2}, the preformed cluster model with Skyrme-like effective interactions \cite{seif1,seif2}.

In 1977 \cite{blo}, the proximity potential was introduced by Blocki and coworkers to estimate the strength of the nuclear interactions between two heavy ions. In general, the nucleus-nucleus interaction potential based on the proximity force theorem can be described as the product of two parts. The one is a factor depending on the shape and geometry of the colliding nuclei, the other is a universal function  $\phi(\xi)$ that depends only on the separation distance $s$ between two colliding nuclei. During recent decades, several modifications have been made in the proximity potential formalism. The main purpose of introducing such modifications is to fix the defects and improve the proximity potential model in the research of fusion reactions and nuclear decays. Accordingly, various versions of the proximity potential have been so far developed  by improving the surface energy coefficient, nuclear radius, surface thickness parameter, and universal function \cite{bl,ngo,gu,zh,my,mol,mol1,dut}. The idea of the universal function is fundamental advantage of proximity potential since it has the simple and accurate formalism. Note that the dimensionless proximity function $\phi(\xi)=\frac{e(\xi b)}{2\gamma}$ is obtained from $e(s)$ by measuring the separation $s$ in units of the surface width $b$ (with $\xi$=$\frac{s}{b}$), and by measuring energy in units of twice the surface energy coefficient $\gamma$ \cite{blo}. Recently, a comprehensive study was conducted on the fusion barriers and cross sections of 15 heavy-ion collision systems with $320 \leq Z_1Z_2\leq 1512$, aiming to better understand the impact of the universal function within the proximity potential formalism on the heavy-ion fusion reactions \cite{Gha}. Three versions of the phenomenological proximity potential (namely Prox. 77, Zhang 2013, and Guo 2013) were selected to calculate the nucleus-nucleus potential. Among the selected potential models, it is shown that Guo 2013 and Prox. 77 yield the lowest and highest fusion barriers for different colliding systems, respectively. Additionally, it has been established that the fusion barrier height systematically increases as the projectile mass number progresses from lighter to heavier nuclei. 

The motivation behind present investigation at the first step is to explore the influence of different universal functions on the alpha-decay process within the framework of the original version of the proximity potential formalism. For this purpose, a collection of 250 alpha emitters with proton number ranging from \textbf{$Z=64$ to 103}  is analyzed. The calculations of the nucleus-nucleus potential is performed using the Prox. 77 along with universal functions proposed in the models gp 77, Guo 2013, Ngo 80, Zhang 2013 and Prox. 2010. We also check the validity of these universal functions to reproduce the experimental data of $\alpha$-decay half-lives of 80 super-heavy nuclei with $Z=104-118$.  By considering the universal function effects, the fusion cross sections at different colliding energies for 10 different $\alpha$-induced reactions  are analyzed. The comparison of fusion cross sections is made between the measured and calculated results for $^{4}$He+$^{40}$Ca, $^{48}$Ti, $^{51}$V, $^{63}$Cu, $^{93}$Nb, $^{162}$Dy, $^{208}$Pb, $^{209}$Bi, $^{235}$U, $^{238}$U fusion systems. We apply the one-dimensional barrier penetration model (1D-BPM) for calculating the theoretical values of fusion cross sections. The structure of the paper is as follows. Section \ref{ana} is divided into three parts: (1) a detailed description of the total interaction potential in the $\alpha$-daughter system within the WKB approximation framework, (2) a brief overview of the various universal functions in the proximity potential model, and (3) the formalism used to calculate the $\alpha$-decay half-lives. Sec. \ref{dis} focuses on a discussion of the results obtained in this study. Finally, the conclusions are presented in Sec. \ref{Con}. 

\section{THEORETICAL FRAMEWORK}\label{ana}
\subsection{Potential formalism}
The total nucleus-nucleus interaction potential $V_{\rm tot}$(r) between the alpha particle and daughter nucleus is calculated as,
\begin{equation} \label{tot}
V_{\rm tot} (r)= V_C (r)+V_{\ell} (r)+V_N (r).
\end{equation}
The Coulomb potential $V_C$(r) is taken as the potential of a uniformly charged sphere with sharp radius $R$, and written as, 
\begin{eqnarray}\label{col}
V_C(r) = Z_{\alpha}Z_de^2\left\{\begin{array}{l}
\frac{1}{r} ~~~~~~~~~~~~~~~~~~~~~~~~~~~r\geq R_C \\
\frac{1}{2R_C}\Bigg[3-\bigg(\frac{r}{R_C}\bigg)^2\Bigg]~~~~ r\leq R_C\\
\end{array} \right.
\end{eqnarray}
where $Z_d$ and $Z_{\alpha}$ denote the atomic number of the daughter nucleus and alpha particle, respectively. $R_C$ is the touching radial separation between the $\alpha$-particle and daughter nucleus. The centrifugal potential $V_{\ell}$ can be calculated as,   
\begin{equation} \label{vl}
V_{\ell} (r)=\frac{\ell(\ell+1)\hbar^2}{2\mu r^2},
\end{equation}
here, $\ell$ and $\mu=\frac{m_{\alpha} m_d}{m_{\alpha}+m_d}$ represent the angular momentum carried away by the emitted alpha particle and the reduced mass of the  $\alpha$-daughter system. We use the original version of the proximity potential formalism for the nuclear potential calculation. According to the proximity 1977 potential \cite{blo} the nuclear potential $V_N$(r) is given as,
\begin{equation} \label{VN}
V_{N}(r)=4\pi b \gamma \overline{R}\phi\big(\xi\big),
\end{equation}
where the surface energy coefficient $\gamma$ is defined as,
\begin{equation} \label{gamma}
\gamma=\gamma_0\Bigg[1-k_s\bigg(\frac{N-Z}{N+Z}\bigg)^2\Bigg],
\end{equation}
here, $\gamma_0$=0.9517 MeV.$\rm fm^{-2}$  and $k_s$=1.7826 \cite{blo} are the surface energy and surface asymmetry constants, respectively. The mean curvature radius (reduced radius) $\overline{R}$ in terms of matter radius $C_i$ is as follows,
\begin{equation} \label{Rbar}
\overline{R}=\frac{C_{\alpha}C_d}{C_{\alpha}+C_d},
\end{equation}
with
\begin{equation} \label{Cbar}
C_i=R_i-\bigg(\frac{b^2}{R_i}\bigg) ~~ (i=\alpha, d),
\end{equation}
where $b$ is the surface thickness parameter with an approximate value of 1 fm. The effective sharp radius $R_i$ of the alpha and daughter nuclei in terms of mass number $A_i$ is, 
\begin{equation} \label{Ri}
R_i=1.28A_i^{1/3}-0.76+0.8A_i^{-1/3} ~~~~   (i=\alpha, d).
\end{equation}
The universal function $\phi(\xi=\frac{s}{b}=\frac{r-C_d-C_\alpha}{b})$ which only depends on separation of the alpha-daughter system is defined as,
\begin{eqnarray}\label{fi}
\phi(\xi)=\left\{\begin{array}{l}-\frac{1}{2}\big(\xi-2.54 \big)^{2}-0.0852\big(\xi-2.54\big)^{3} 
	\\$ for$~~ \xi \leq 1.2511 \\
\ -3.437~{\rm  exp}(-\frac{\xi}{0.75})
\\
$ for$~~ \xi\geq 1.2511\\
\end{array}.\right.
\end{eqnarray}

\subsection{The various universal functions} 
In this study, beyond the formulation presented in the Prox. 77 model, several alternative forms of universal functions are used to calculate the nucleus-nucleus potential, as detailed in the following sections,

\subsubsection{The generalized proximity potential 1977 (gp 77)}
The universal function $\phi(\xi)$ for the generalized proximity potential 1977 (gp 77) \cite{sant} is calculated by,
\begin{eqnarray}\label{gp77}
\phi(\xi)=\left\{\begin {array}{l}
\ -1.7817+0.927\xi+0.0169\xi^{2}-0.05148\xi^{3}
\\$ for$ ~~ 0.0 \leq \xi \leq 1.9475\\
\ -4.41~{\rm exp}(-\frac{\xi}{0.7176})
\\$ for$ ~~ \xi> 1.9475\\
\end{array}\right.
\end{eqnarray}

\subsubsection{Ngo 80}
For Ngo 80 \cite{ngo}, the universal function is determined as follows, 
\begin{eqnarray}\label{Ngo 80}
\phi(s)=\left\{\begin{array}{l}
\ -33+5.4(s-s_0)^{2}  ~~~~~~~~~~ $for$ ~~s < s_0\\
\ -33~{\rm exp}\big[-\frac{1}{5}(s-s_0)^{2}\big]~~~~$for$ ~~ s\geq s_0\\
\end{array}, \right.
\end{eqnarray}
 here, $s_0=-1.6$ fm. 
\subsubsection{Guo 2013 and Zhang 2013}
In Guo 2013 \cite{gu} and Zhang 2013 \cite{zh} potential models, the universal function has the following form, 
\begin{equation} \label{GZ}
	\phi(s)=\frac{p_1}{1+{\rm exp}\bigg(\frac{s+p_2}{p_3}\bigg)},
\end{equation}
where the constant values of $(p_1, p_2, p_3)$ for Guo 2013 (Zhang 2013) proximity potential is reported as -17.72, 1.30, 0.854 (-7.65, 1.02, 0.89). 

\subsubsection{Prox. 2010}
In Prox. 2010 model \cite{dut}, the universal function is given as,
\begin{eqnarray}\label{Prox2010}
\phi(\xi)=\left\{\begin{array}{lr}
\ -1.7818+0.9270\xi+0.143\xi^{3}-0.09\xi^{3}  \\ $for$~~~
\xi\leq 0.0\\
\ -1.7817+0.9270\xi+0.01696\xi^{2}-0.05148\xi^{3} \\$for$~~~
 0.0 \leq \xi \leq 1.9475\\
\ -4.41{\rm exp}(-\frac{\xi}{0.7176}) \\$for$~~~
 \xi \geq 1.9475\\
\end{array}\right. 
\end{eqnarray}

\subsection{Alpha decay half-life formalism}
The alpha decay half-life can be calculated by, 
\begin{equation} \label{T12}
T_{1/2}=\frac{\rm ln 2}{\nu P_0 P_\alpha }.
\end{equation}
The alpha decay penetration probability $P_\alpha$ through the interaction potential is determined based on the one dimensional WKB approximation as,
\begin{equation} \label{pene}
P_\alpha=\textmd{exp}\bigg[-\frac{2}{\hbar}\int_{R_a}^{R_b}\sqrt{2\mu
(V_{\rm tot}(r)-Q_{\alpha})}dr\bigg],
\end{equation}
where $V_{\rm tot}$ and $Q_{\alpha}$ are the total interaction potential and the released energy of the emitted $\alpha$-particle, respectively. Moreover, $R_a$ and $R_b$ refer to the physical turning points and are given by,
\begin{equation} \label{cond}
	V_{\rm tot}(R_a)=Q_{\alpha}=V_{\rm tot}(R_b).
\end{equation}
The preformation factor $P_{0}$ of alpha particle was obtained as 0.43 for even-even, 0.35 for odd-A and 0.18 for odd-odd nuclei \cite{Xu}.
Here $\nu = \frac{\omega}{2\pi}=\frac{2E_{\nu}}{h},$ is known as the assault frequency (refers to the number of alpha particle attacks on the barrier per second), where $h$ and $E_{\nu}$ are Planck constant and the empirical vibrational energy, respectively \cite{Roy}. 

\section{RESULTS AND DISCUSSION}\label{dis}
During recent years, the proximity potential model was generally used for various studies on the alpha-decay. Among the different versions of proximity potentials, it is shown that Prox. 77 can work well in the connection with the alpha-decay studies with low deviation, see for example Refs. \cite{sant1,Zhan1}.  Accordingly, this potential model is adopted in the present work for estimating the nuclear interactions between $\alpha$-particle and daughter nucleus. We are interested in investigating the effect of the universal function of the proximity potential model on the alpha-decay process. To reach this goal, the $\alpha$-decay half-lives for the emission of the alpha particle from 250 parent nuclei in the atomic number range of $Z = 64-103$ have been analyzed using Prox. 77 along with five different universal functions $\phi^{gp~77}$, $\phi^{Guo~2013}$, $\phi^{Ngo~80}$,  $\phi^{Zhang~2013}$, and $\phi^{Prox.~2010}$. In this study, the results of the modified versions of proximity 77 are marked as MP-I, MP-II, MP-III, MP-IV, MP-V, respectively.

\begin{figure*}[htp]
	\vspace{-5pt}
	\centering
	\includegraphics[width=0.70\textwidth]{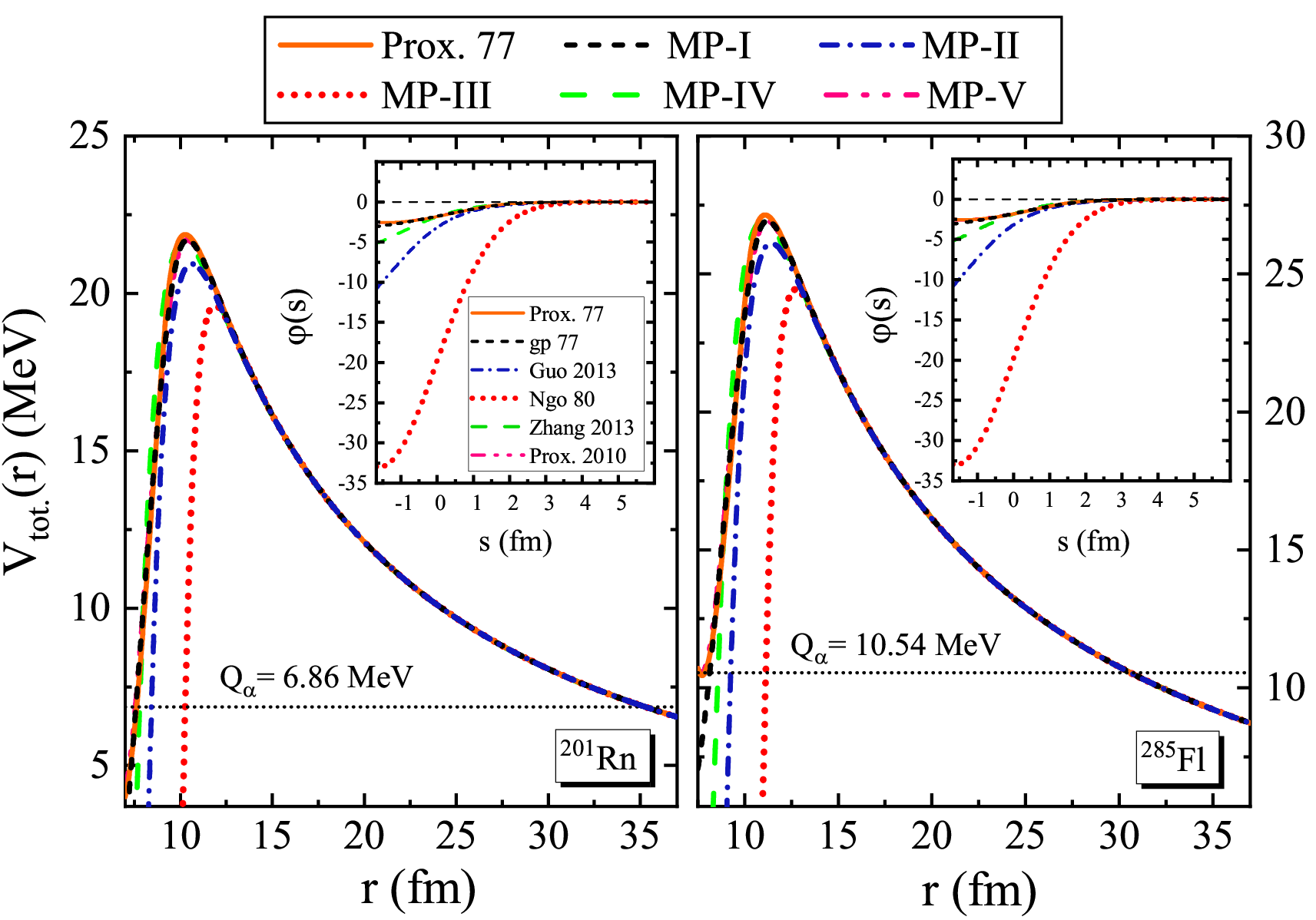}
	\vspace{-5pt}
	\caption{\footnotesize (Colored online) The total interaction potentials versus the radial distance $r$ (in fm) using Prox. 77 with different universal functions for alpha-decay from $^{201}$Rn (left panel) and $^{285}$Fl (right panel)  parent nuclei, as an example. The variation of the universal functions with separation distance $s$ (in fm) are given in the insert.}
	\label{Vtot}
\end{figure*}

In Fig. \ref{Vtot} (left panel), we compare the radial behavior of total interaction potential constructed from the original version of the proximity potential with those obtained from its modified forms for alpha-decay of $^{201}$Rn parent nucleus with released energy $Q_\alpha= 6.86$ MeV as an example. To further study the role of universal function in the emitted alpha-daughter nucleus interaction potential distribution, the different shapes of $\phi(s)$ are plotted as a function of short separation distance $s$ in the insert of the figure. Note that the calculations of $\phi(s)$ have been extended from positive to negative values of $s$. One can see that the difference of the calculated universal functions based on the Prox. 77, gp 77, and Prox. 2010 models is small. While, Fig. \ref{Vtot} shows that the universal functions of Ngo 80, Guo 2013, and Zhang 2013 depend strongly on the separation distance between two nuclei. What needs to be emphasized is that the deviation of the results obtained from these three models increases when $s$ becomes more negative. In contrast, the various formulae of universal function tend to be the same when the values of the short separation distance increases in the positive $s$ region. Since the nuclear proximity potential $V_N$(r) is directly dependent on $\phi(s)$, one can conclude that the potentials derived from the Prox. 77, MP-I, and MP-V models are significantly less attractive than those calculated using the MP-II, MP-III, and MP-IV models. Accordingly, it is indicated that the interaction potentials calculated using the universal functions of Ngo 80, Guo 2013 and Zhang 2013 models provide the lowest barrier heights, respectively. The values of barrier height $V_B$ (in MeV) using the different considered models are tabulated in Table \ref{height} for alpha-decay of $^{201}$Rn parent nucleus. 

\begin{figure}[htp]
	\vspace{-5pt}
	\centering
	\includegraphics[width=0.75\textwidth]{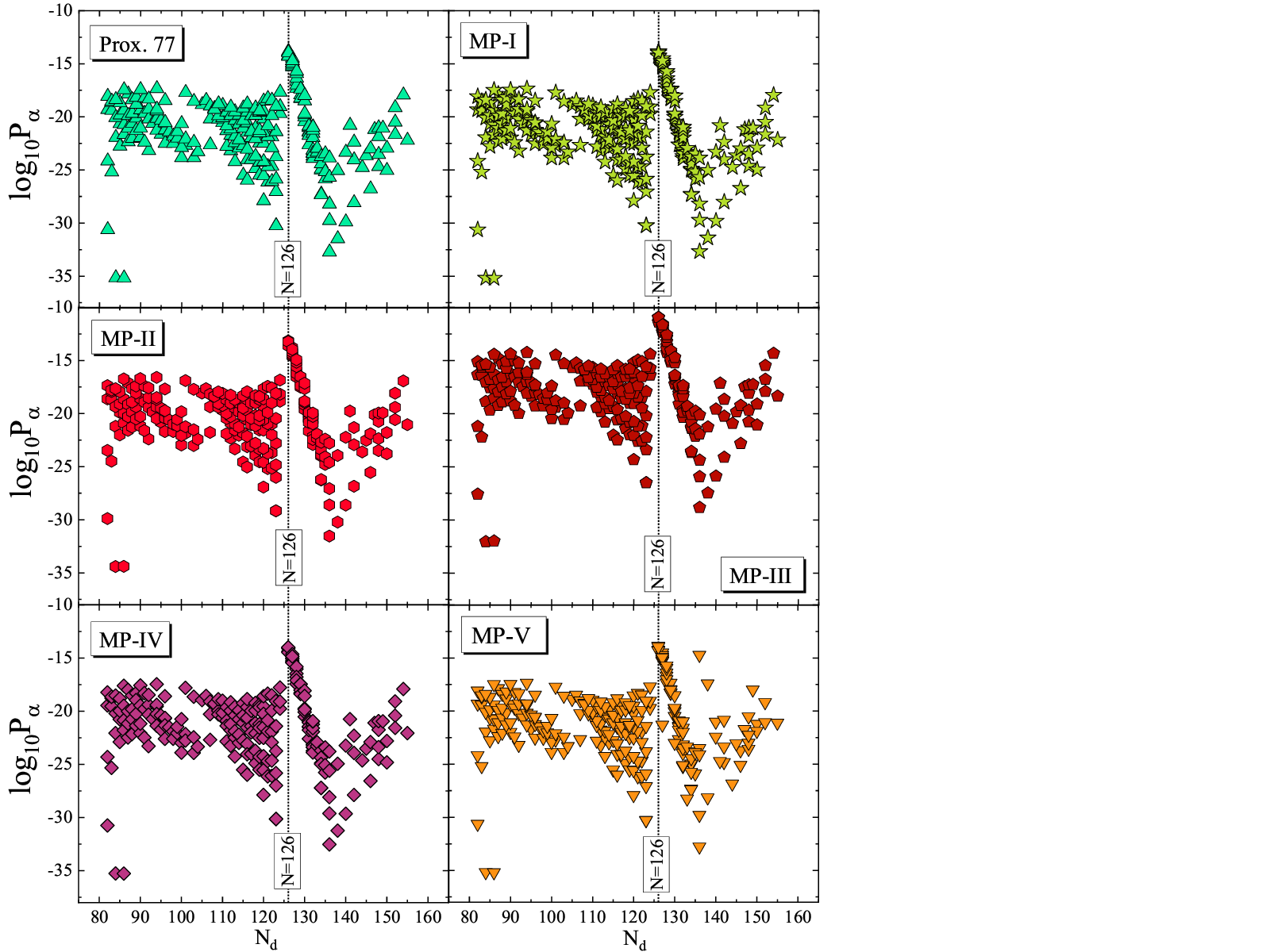}
	\vspace{-10pt}
	\caption{\footnotesize (Colored online) The behavior of the logarithmic values of the penetration probability using the original and modified forms of the Prox. 77 potential in terms of the neutron number of the daughter nuclei $N_d$. The vertical dotted lines show the magic number $N_d=126.$}
	\label{logP}
\end{figure}

\begin{figure}[htp]
	\vspace{-5pt}
	\centering
	\includegraphics[width=0.75\textwidth]{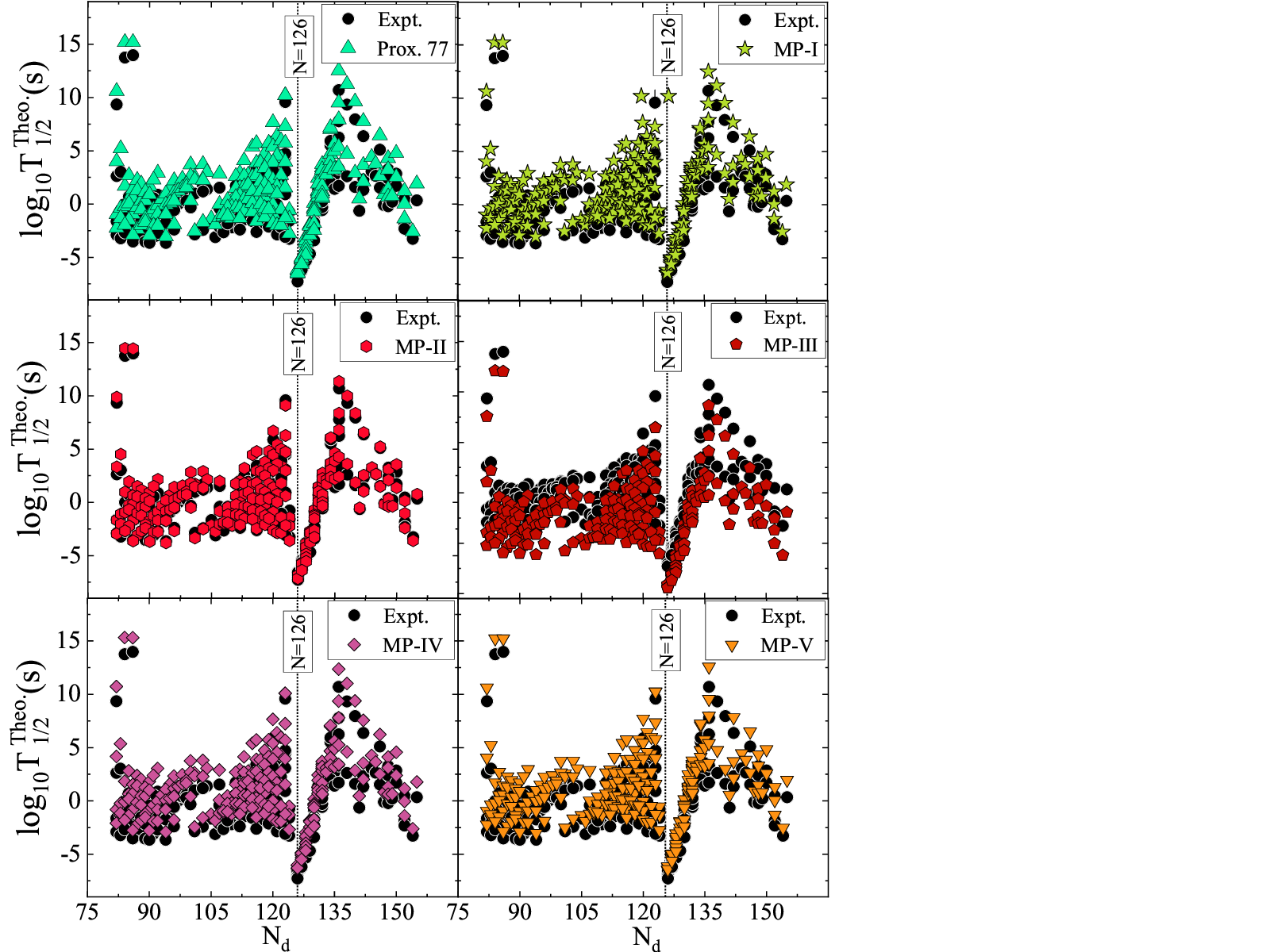}
	\vspace{-10pt}
	\caption{\footnotesize (Colored online) The behavior of the logarithmic values of the experimental half-life data (gray circle) and the calculated half-lives (colored circle) using Prox. 77 model with different universal functions versus the neutron number of the daughter nuclei $N_d$.  The vertical dotted lines show the magic number $N_d=126.$}
	\label{logT}
\end{figure}

It is expected that the observed variations of the interaction potential between the emitted $\alpha$-particle and the daughter nucleus affect the alpha-decay penetration probability through the potential barrier and result in the alpha-decay half-lives. Fig. \ref{logP} illustrates the logarithmic behavior of the penetration probability $P_{\alpha}$ as a function of the neutron number of the daughter nucleus $N_{d}$. The calculated values of $P_{\alpha}$ present a vibrations pattern as the neutron number approaches the magic number $N_d=126$, marked with a dotted line in the figure. It is worth noting that the shell effect in alpha decay is closely related to the released energy $Q_{\alpha}$. In fact, this effect enhances the stability of the daughter nucleus near magic numbers, leading to an increase in the $Q_{\alpha}$-value, a higher penetration probability, and consequently, a reduction in the alpha-decay half-life. This result is clearly observed in Fig. \ref{logT}. In this figure, the logarithmic behavior of the calculated half-lives in terms of the neutron number of the daughter nucleus $N_d$ is shown for all selected parent nuclei. It is observed that the effects of the closed shell at the magic number $N_d=126$ can well reproduce by different potential models. In addition, one can find that the alpha radioactivity half-lives using the MP-III model are smaller than the other ones.
\begin{table*}[htp]
	\caption{The calculated barrier heights $V_B$ (in MeV) using the original and modified forms of the Prox. 77 potential for $\alpha$-decay $^{201}$Rn and $^{285}$Fl parent nuclei, for example.} \centering
	\begin{tabular}{ccccccccc}
		\\
		\hline\hline
& \multicolumn {6}{c} {Proximity potential}\\
  \cline{2-7} 
		Parent nucleus &  MP-I &~~ MP-II &~~ MP-III  &~~ MP-IV &~~ MP-V &~~ Prox. 77\\
		\hline
		 $^{201}$Rn &~~ 21.69 &~~ 20.93 &~~ 19.61&~~ 21.61 &~~ 21.69 &~~ 21.86 \\
\hline
$^{285}$Fl  &~~ 26.92 &~~ 26.10 &~~ 24.44  &~~ 26.92 &~~ 26.92 &~~ 27.13\\
\hline\hline
		\end{tabular}
	\label{height}
\end{table*}

\begin{figure}[htp]
	\vspace{-5pt}
	\centering
	\includegraphics[width=0.62\textwidth]{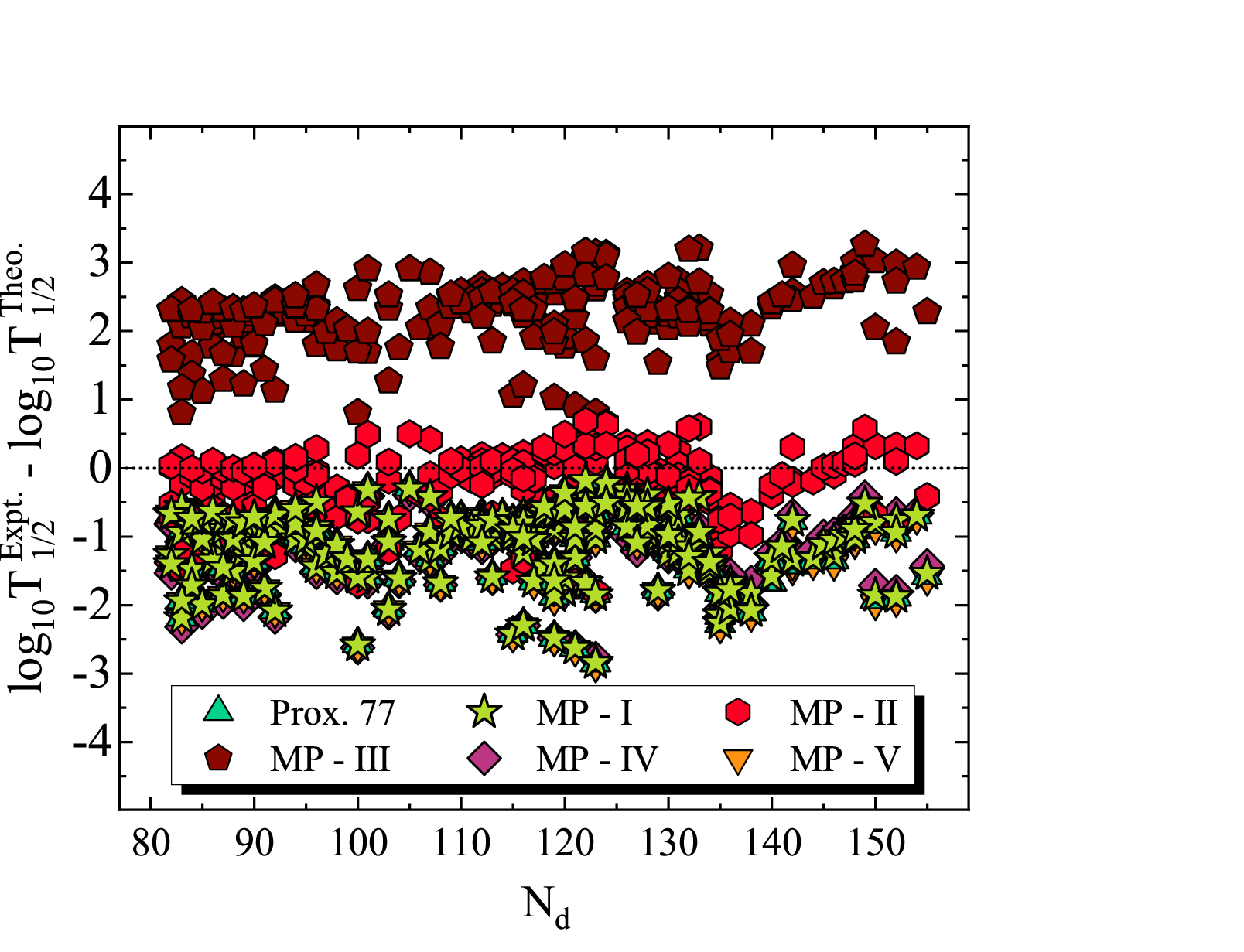}
	\vspace{-5pt}
	\caption{\footnotesize (Colored online) The difference between the predicted alpha half-lives and the experimental data using the original and modified forms of the Prox. 77 potential against the neutron number of daughter nuclei $N_d$. }
	\label{deltaT}
\end{figure}

The logarithmic difference between the calculated $\alpha$-decay half-lives and the experimental data is presented in Fig. \ref{deltaT}. It is clear that the half-lives calculated using the MP-III model exhibit significant deviations from the experimental data. In contrast, the results from the MP-I, MP-IV and MP-V models show similar trends, in agreement with the results obtained from the original Prox. 77 model. The half-lives calculated with the MP-II model show the closest agreement with the experimental data, demonstrating a smaller deviation than the other models. For a more detailed examination, the standard deviation $\sigma$ of the calculated half-life values from the corresponding experimental data is obtained through the following equation,
\begin{equation} \label{sigma1}
	\sigma=\sqrt{\frac{1}{N}\sum_{i=1}^{N}\bigg[\rm{log}_{10}\bigg(T^{Theo.}_{1/2i}\bigg)-\rm{log}_{10}\bigg(T^{Expt.}_{1/2i}\bigg)\bigg]^2},
\end{equation}
where $N$ is the number of parent nuclei used for evaluation of the $\sigma$ value. The $\sigma$ values for the studied models are presented in the Table. \ref{sigma}. Analysis of the data reveals that the MP-II model provides the most accurate estimation of $\alpha$-decay half-lives with $\sigma= 0.49$. This means that the universal function of the Guo 2013 potential model provides a significant improvement over the original proximity potential 1977 (with $\sigma= 1.17$). Moreover, it is evident that the MP-III model unable to accurately reproduce the experimental half-life data. Therefore, one can conclude that the Prox. 77 potential along with the $\phi^{Ngo~80}$ is not suitable to deal with the alpha radioactivity.

\begin{table*}[htp]
	\caption{The obtained standard deviations $\sigma$ of the calculated alpha decay half-life values for original and modified forms of Prox. 77 using various universal functions in comparison with the experimental data.} \centering
	\begin{tabular}{ccccccc}
		\hline\hline   Model &~~ MP-I &~~ MP-II &~~ MP-III&~~ MP-IV &~~ MP-V&~~ Prox. 77 \\
		\hline 
		$\sigma$ &~~ 1.18 &~~ 0.49 &~~ 2.33&~~ 1.20 &~~ 1.19 &~~ 1.17 \\
	\hline
	\end{tabular}
	\label{sigma}
\end{table*}

\subsubsection{The Geiger-Nuttall law}
\begin{figure}[htp]
	\vspace{-5pt}
	\centering
	\includegraphics[width=0.77\textwidth]{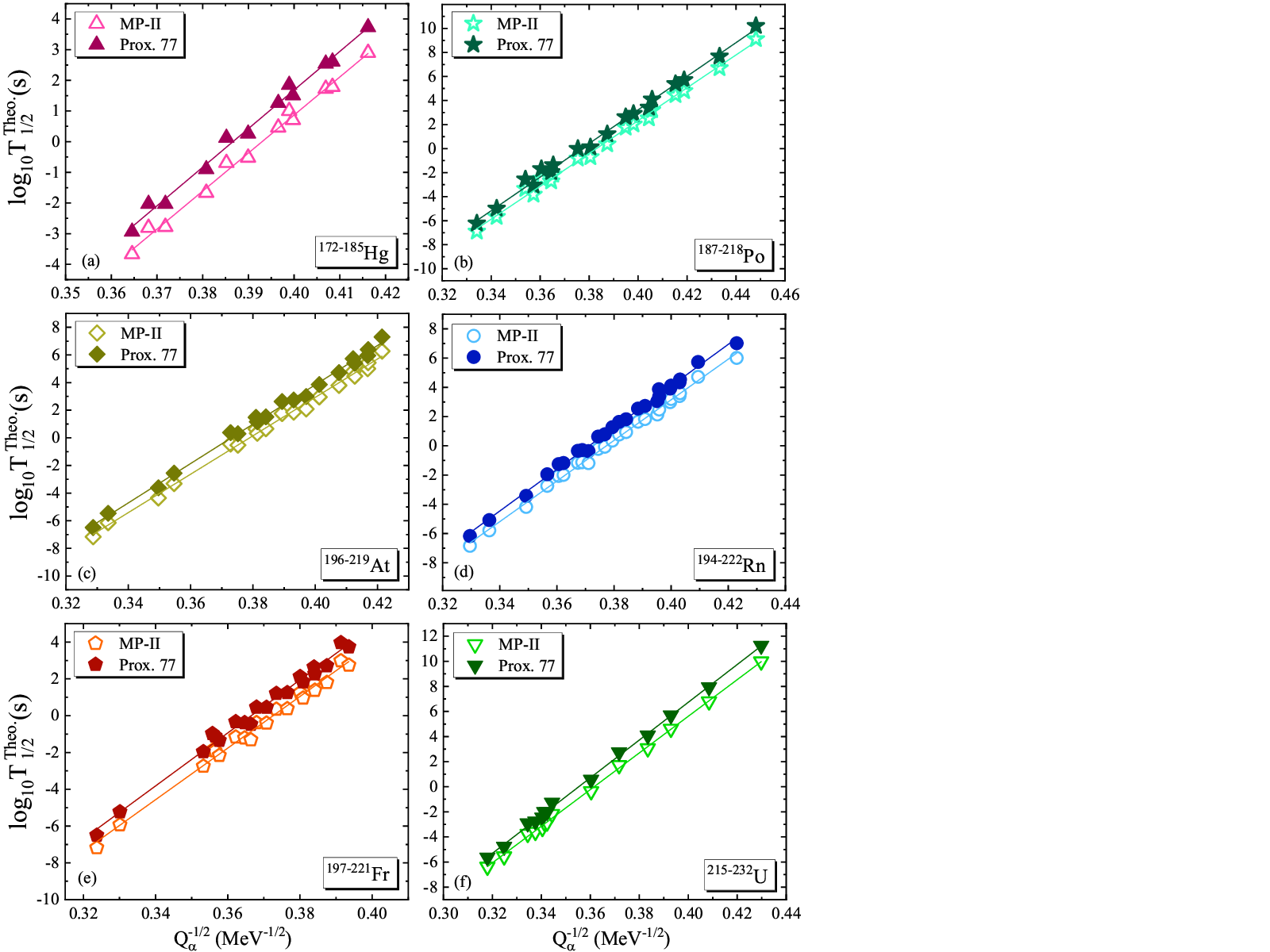}
	\vspace{-5pt}
	\caption{\footnotesize (Colored online) Geiger-Nuttall plot of $\ log _{10} (T_{1/2})$ values versus $\ Q^{-1/2}_{\alpha}$ for the emission of $\alpha$-particle from (a) $^{172-185}$Hg, (b) $^{187-218}$Po, (c)
		$^{196-219}$At, (d) $ ^{194-222}$Rn, (e) $^{197-221}$Fr and (f) $^{215-232}$U parent nuclei using the MP-II (solid circles) and Prox. 77 (up-pointing triangles) potential models. $\ T_{1/2}$ is in seconds. }
	\label{GN}
\end{figure}

The Geiger-Nuttall (GN) law is a well-established empirical relation in nuclear physics that describes the correlation between the half-life of $\alpha$-decay and the energy released by the emitted alpha particle \cite{Geiger}. It states that the logarithm of the alpha-decay half-life is linearly related to the inverse square root of the released energy $Q_{\alpha}$. Accordingly, this study is conducted to investigate the validity of the GN law in the behavior of calculated alpha-decay half-lives for a set of selected heavy nuclei based on the MP-II potential model. The GN plots for the isotopes of $^{172-185}$Hg (with Z=80), $^{187-218}$Po (with Z=84), $^{196-219}$At (with Z=85), $ ^{194-222-185}$Rn (with Z=86), $^{197-221}$Fr (with Z=87) and $^{215-232}$U (with Z=92) are presented in Fig. \ref{GN}, for example. It is evident from the figure that, for various isotopic groups, the logarithm of the half-life values exhibits a clear linear correlation with the decay energy $Q^{-1/2}_{\alpha}$. Therefore, the linearity observed in the GN plots supports the validity of the MP-II potential model. For comparison, in Fig. \ref{GN}, the GN plots for $^4$He emission from various isotopic chains are also presented based on the Prox. 77 potential model. One can see that the half-lives calculated by the Prox. 77 are greater than those obtained from the MP-II potential model. A linear trend is observed in all cases, with variations in slope and intercept.

\subsubsection{Super-heavy nuclei}
\begin{figure}[htp]
	\vspace{-5pt}
	\centering
	\includegraphics[width=0.74\textwidth]{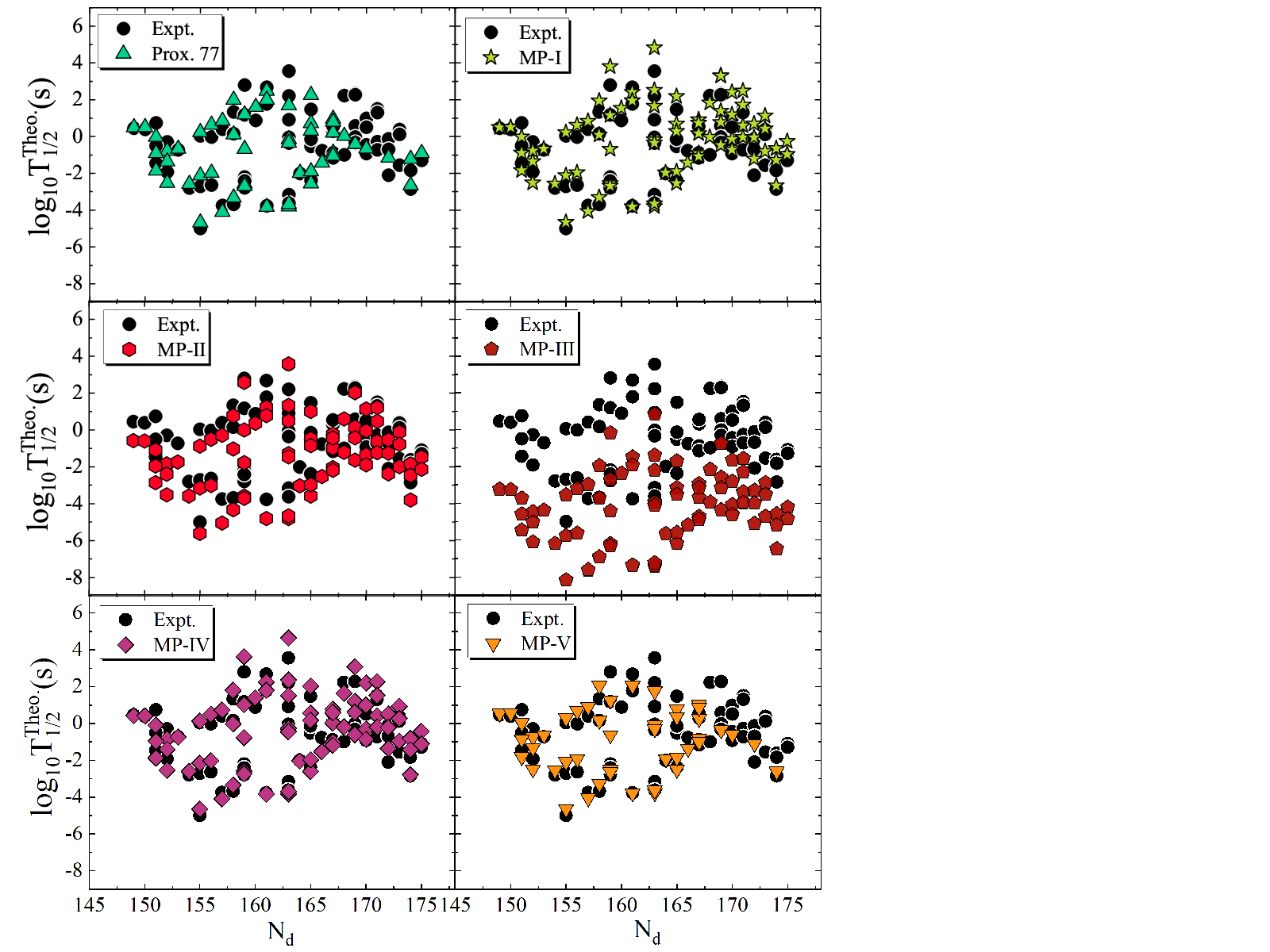}
	\vspace{-10pt}
	\caption{\footnotesize (Colored online) The behavior of the logarithmic values of the experimental half-life data (black circle) and the calculated half-lives (colored hexagonal) of super-heavy nuclei using original and modified forms of Prox. 77 in terms of the neutron number of the daughter nuclei $N_d$.}
	\label{logTSHN}
\end{figure}

During recent decades, the study of superheavy nuclei (SHN) has become an interesting subject in modern nuclear physics research, both theoretically and experimentally. As earlier stated, alpha-decay is one of the most important decay modes for such parent nuclei. Therefore, it can be of significant interest to study the role of universal function of the proximity potential in the $\alpha$-decay half-lives of SHN regions. For this purpose, the $\alpha$-decay half-lives of 80 super-heavy nuclei with $Z=104-118$ are investigated by employing the original and modified forms of the proximity potential formalisms. Fig. \ref{Vtot} (right panel) shows the radial behavior of the different universal functions and their effects on the total interaction potentials as a function of radial distance $r$ (fm) for alpha decay of $^{285}$Fl parent nucleus with released energy $Q_{\alpha}=10.54$ MeV. We observe a similar patterns for the heavy and super-heavy nuclei in the left and right panels of the figure. The only difference is that the potential barriers generated in the SHN region are higher compared to the heavy parent nuclei. This result can be interpreted by the higher Coulomb repulsion for nuclei with higher atomic number. To enable a detailed comparison, we present the barrier heights $V_{B}$ (in MeV) obtained from the various studied models in Table \ref{height} for $^{285}$Fl parent nucleus. In Fig. \ref{logTSHN}, a comparison between the logarithmic values of alpha radioactivity half-lives of calculations and experimental data have been done as a function of the neutron number of different daughter nuclei. Note that for Prox. 77 and MP-V potential models, the calculations are limited to fewer parent nucleus because there is no inner turning point in some nuclei. It can be seen that the calculated half-lives using MP-III model have the lowest values among the other 5 versions of the proximity potential formalisms.

\begin{figure}[htp]
	\vspace{-5pt}
	\centering
	\includegraphics[width=0.60\textwidth]{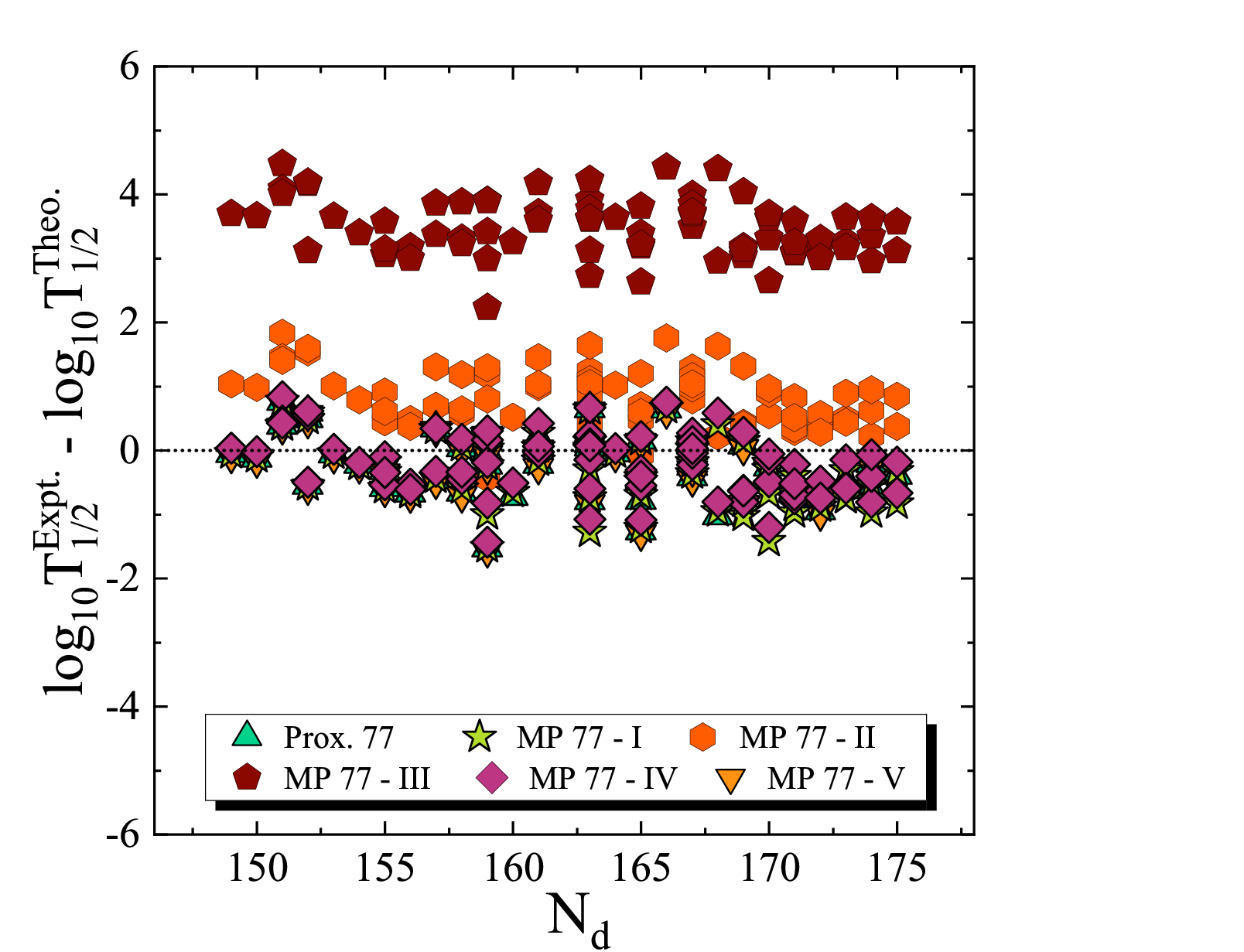}
	\vspace{-5pt}
	\caption{\footnotesize (Colored online) The difference between the predicted alpha half-lives and the experimental data of the superheavy nuclei using the original and modified forms of the Prox. 77 potential against the neutron number of daughter nuclei $N_d$. }
	\label{delTSHN}
\end{figure}

The logarithmic differences between the theoretical and experimental $\alpha$-decay half-lives is displayed in Fig. \ref{delTSHN} for the SHN of interest. It is clear from the figure that the MP-III potential model shows a considerable deviation with respect to available experimental data. Whereas, the results of other models have a behavior close to the experimental half-lives.To gain further insight, the standard deviation $\sigma$ of logarithm values of the theoretical half-lives resulting from the original and modified forms of the proximity potential in comparison with the corresponding experimental data is calculated for 80 SHN thorough Eq.(\ref{sigma1}). One can see that the MP-IV (with $\sigma$=0.47) and MP-I (with $\sigma$=0.50) are appropriate nuclear potentials that can predict the alpha decay half-lives of super-heavy nuclei with least standard deviation, respectively. In addition, it is found that MP-III is not suitable to deal with the alpha radioactivity of superheavy nuclei.

\begin{table*}[htp]
	\caption{The obtained standard deviations $\sigma$ of the calculated alpha decay half-life values of super-heavy nuclei for original and modified forms of Prox. 77 using various universal functions in comparison with the experimental data.The numbers in parentheses are the number of nuclei under consideration.} \centering
	\begin{tabular}{ccccccccc}
		\hline\hline   Model &~~ MP-I &~~ MP-II&~~ MP-III&~~ MP-IV&~~ MP-V&~~ Prox. 77   \\
		\hline 
		$\sigma$ &~~ 0.50 (80) &~~ 1.02 (80)&~~ 3.59 (80)&~~ 0.47 (80)&~~ 0.53 (51)&~~ 0.51 (58)   \\
		\hline
	\end{tabular}
	\label{sigma2}
\end{table*}

\subsubsection{$\alpha$-induced fusion reactions}

\begin{figure*}[htp]
	\vspace{-5pt}
	\centering
	\includegraphics[width=0.85\textwidth]{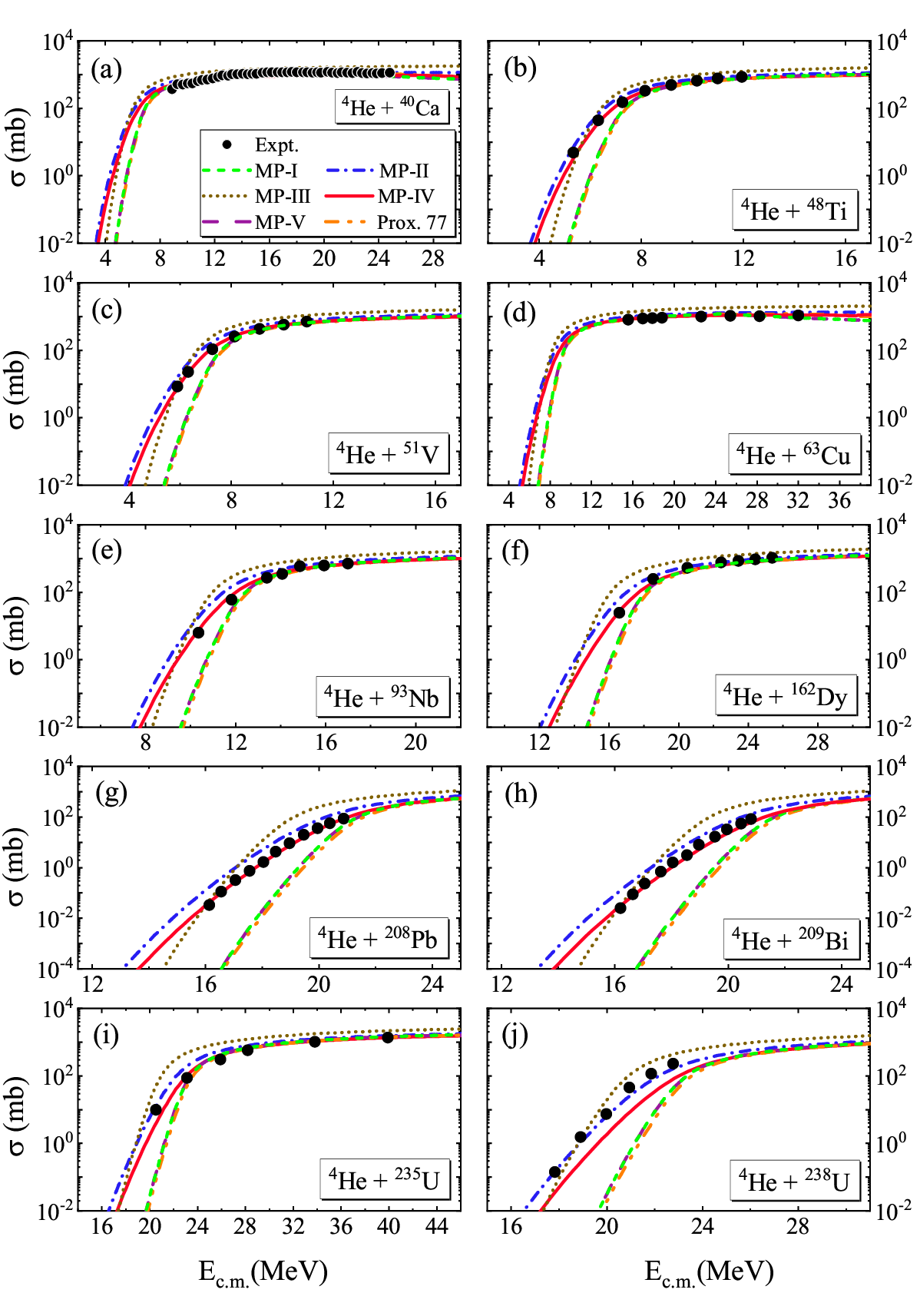}
	\vspace{-5pt}
	\caption{\footnotesize (Colored online) The energy-dependent behavior of the experimental fusion cross sections for the reactions (a) $\alpha$ + $^{40}$Ca, (b) $\alpha$ + $^{48}$Ti, (c) $\alpha$ + $^{51}$V, (d) $\alpha$ + $^{63}$Cu, (e) $\alpha$ + $^{93}$Nb, (f) $\alpha$ + $^{162}$Dy, (g) $\alpha$ + $^{208}$Pb, (h) $\alpha$ + $^{209}$Bi, (i) $\alpha$ + $^{235}$U and (j) $\alpha$ + $^{238}$U compared with the Prox. 77 potential model with different universal functions. The experimental data are taken from Refs. \cite{HeCa,HeTi,HeCu,HeNb,HeDy,HeU5, HeU8, HeU}}
	\label{cross}
\end{figure*}

Alpha-induced fusion reactions involve the interaction of an alpha particle, leading to the formation of a heavier compound nucleus. These reactions play a significant role in nuclear astrophysics, particularly in stellar nucleonsynthesis \cite{dre,ozk}. In contrast to alpha-decay, where an unstable nucleus spontaneously emits an alpha particle to become more stable, $\alpha$-induced fusion is a non-spontaneous, energy-dependent process requiring high kinetic energy to overcome the Coulomb barrier between nuclei. On the other hand, the fusion reaction between the $\alpha$-particle and the target nucleus occurs in the opposite direction to the $\alpha$-decay. The study of such fusion reactions often involves detailed cross-section measurements and theoretical models to understand the interaction between two colliding nuclei. Under these conditions, one can expect that the nuclear potential plays an indispensable role in the fusion reactions. Hence, it can be interesting to explore the validity of the modified proximity potentials for reproducing the energy-dependent behavior of the fusion cross sections in $\alpha$-induced reactions. The present study focuses on $\alpha$-induced reactions on intermediate mass and heavy nuclei, including $^{4}$He+$^{40}$Ca, $^{48}$Ti, $^{51}$V, $^{63}$Cu, $^{93}$Nb, $^{162}$Dy, $^{208}$Pb, $^{209}$Bi, $^{235}$U, $^{238}$U fusion systems. In Fig. \ref{cross}, the experimental fusion cross sections for the selected reactions are compared with those obtained theoretically thorough the Prox. 77 potential model with different universal functions. The simple approach of one-dimensional barrier penetration model has been employed for calculation of cross sections where both the projectile and the target are assumed to be structureless. From the figure, one can found that the calculated cross sections values using the MP-IV model is very close to the corresponding experimental data for the reactions involving light and medium nuclei (a-h panels). Fig. \ref{cross} indicates that the MP-II model provides a more accurate description of the experimental cross-section data for reaction (j) $^{4}$He+$^{238}$U at sub-barrier energies. A comparable pattern is also evident in the case of reaction (i)$^{4}$He+$^{235}$U. These results corroborate the findings presented in Ref. \cite{Gha}.
	
\section{Conclusions}\label{Con}
To investigate the impact of universal functions on the alpha decay process, a comparative study is conducted using the universal functions of the gp 77, Ngo 80, Guo 2013, Zhang 2013, and Prox. 2010 models. The selected universal functions are incorporated into the original version of the proximity potential (Prox. 1977), and subsequently used to calculate the theoretical $\alpha$-decay half-lives of 250 parent nuclei with atomic numbers $Z = 64-103$, within the framework of the WKB approximation. The theoretical results are evaluated against the corresponding experimental data using the root-mean-square deviation ($\sigma$). Additionally, to enable a more effective comparison, the half-life values for selected $\alpha$-decays are also calculated using the original version of proximity potential model. The analysis revealed that the universal function $\phi$(s) of the Guo 2013 model, when incorporated into the proximity theory (MP-II), yields the most reliable predictions for $\alpha$-decay half-lives. The MP-II model overcomes the limitations of the original Prox. 77 potential, leading to a notable decrease in the standard deviation ($\sigma$) from 1.17 to 0.49. It should be noted that the MP-I, MP -IV, and MP-V models yield the standard deviation values close to those obtained for the original version of the proximity model. However, the MP-III model with $\sigma$ = 2.33 can not be suitable for predicting half-lives in heavy nuclei. Encouraged by this, we examine the validity of the GN law in the behavior of calculated alpha-decay half-lives for the various isotopic groups based on the MP-II potential model. Our detailed investigation reveals that the $\alpha$-decay half-life values for the ground-state to ground-state transitions follow a regular linear trend as a function of $\ Q^{-1/2}_{\alpha}$ ($MeV^{-1/2}$). It indicates that the GN law can be confirmed in various isotopic groups within the framework of the MP-II potential model. we performed a comparative study of 6 versions proximity potential formalisms applied to alpha radioactivity half-lives of 80 super-heavy nuclei (SHN) with $Z=104-118$. The comparison between the calculated half-lives and the corresponding experimental data reveals that the standard deviation of the decimal logarithm of the half-lives based on the MP-IV potential model (with $\sigma$ = 0.47) has a smaller value than the other versions of the proximity potential formalisms. Whereas, MP-III model (with $\sigma$ = 3.59) shows a largest deviation with the experimental data. Finally, the behavior of fusion cross sections for alpha-capture reactions on $^{40}$Ca, $^{48}$Ti, $^{51}$V, $^{63}$Cu, $^{93}$Nb, $^{162}$Dy, $^{208}$Pb and $^{209}$Bi, $^{235}$U, and $^{238}$U nuclei have been evaluated using the original and modified versions of the nuclear proximity potentials. The obtained results show that the fusion cross sections based on the MP-IV potential model are consistent with the corresponding experimental data.  It can be stated that the MP-II model performs better in reproducing the experimental cross-section data for $^{4}$He+$^{235}$U, $^{238}$U reactions at sub-barrier energies.

\end{document}